%% file: acl_latex.tex
\pdfoutput=1

\documentclass[11pt]{article}

\usepackage[final]{acl}

\usepackage{times}
\usepackage{latexsym}

\usepackage[T1]{fontenc}
\usepackage{mathrsfs}

\usepackage[utf8]{inputenc}

\usepackage{microtype}

\usepackage{inconsolata}

\usepackage[ruled,vlined]{algorithm2e}
\usepackage{graphicx}
\usepackage{multirow}
\usepackage{dirtytalk}
\usepackage{amsfonts}
\usepackage{amsmath}
\usepackage{subcaption}
\usepackage{booktabs}
\usepackage{colortbl}
\title{Progressive Facial Granularity Aggregation with Bilateral Attribute-based Enhancement for Face-to-Speech Synthesis}

\author{Yejin Jeon$^1$, Youngjae Kim$^1$, Jihyun Lee$^1$, Hyounghun Kim$^{1,2}$, Gary Geunbae Lee$^{1,2}$ \\
  $^1$Graduate School of Artificial Intelligence, POSTECH, Republic of Korea\\
  $^2$Department of Computer Science and Engineering, POSTECH, Republic of Korea\\
  \texttt{\{jeonyj0612, yj122198, jihyunlee, h.kim, gblee\}@postech.ac.kr} \\
}

\begin{document}
\maketitle
\begin{abstract}
For individuals who have experienced traumatic events such as strokes, speech may no longer be a viable means of communication. While text-to-speech (TTS) can be used as a communication aid since it generates synthetic speech, it fails to preserve the user’s own voice. As such, face-to-voice (FTV) synthesis, which derives corresponding voices from facial images, provides a promising alternative. However, existing methods rely on pre-trained visual encoders, and finetune them to align with speech embeddings, which strips fine-grained information from facial inputs such as gender or ethnicity, despite their known correlation with vocal traits. Moreover, these pipelines are multi-stage, which requires separate training of multiple components, thus leading to training inefficiency. To address these limitations, we utilize fine-grained facial attribute modeling by decomposing facial images into non-overlapping segments and progressively integrating them into a multi-granular representation. This representation is further refined through multi-task learning of speaker attributes such as gender and ethnicity at both the visual and acoustic domains. Moreover, to improve alignment robustness, we adopt a multi-view training strategy by pairing various visual perspectives of a speaker in terms of different angles and lighting conditions, with identical speech recordings. Extensive subjective and objective evaluations confirm that our approach substantially enhances face-voice congruence and synthesis stability.
\end{abstract}


\input{introduction}

\input{relatedwork}
\input{method}

\input{experiments}
\input{results}
\input{conclusion}
\input{limitation}

\bibliography{custom}

\appendix

\begin{algorithm}[h]
\caption{Face Extraction and Speaker-Wise Pseudo-Label Assignment}
\label{alg:pseudo_labeling}
\SetAlgoLined
\SetKwInOut{Input}{Input}
\SetKwInOut{Output}{Output}

\Input{Video $\mathcal{V}$}
\Output{Face images $\mathcal{I} = \{I_1, \dots, I_5\}$, Attribute label $\mathcal{A}$}

\BlankLine
\textbf{Step 1: Frame Sampling} \\
$\mathcal{F} \gets \text{s3fd}(\mathcal{V})$ \\
$\mathcal{F} \gets \text{shuffle}(\mathcal{F}) $\\
$\mathcal{I} \gets \mathcal{F}[:5]$ \\

\BlankLine
\textbf{Step 2: Pseudo-Label Generation} \\
$\mathcal{A} \gets \textsc{Null} $\\

\ForEach{$I_k \in \mathcal{I}$}{
    \If{$\mathcal{A} = \textsc{Null}$}{
        $(\text{attr}_1, \text{attr}_2) \gets \text{DeepFace}(I_k)$ \;
        $\mathcal{A} \gets [\text{attr}_1, \text{attr}_2]$ \;
    }
}

\BlankLine
\Return $\mathcal{I}$, $\mathcal{A}$
\end{algorithm}

\section{Dataset Processing}
\label{sec:dataset_processing}
The video from a speaker can be viewed as a sequence of individual frames, each representing a still image. From these images, the face is extracted using the s3fd model. Five frames are randomly selected for further processing. After selecting the five images, one image from each speaker is annotated using the DeepFace pretrained model, as described in Algorithm 1. The remaining four images inherit the annotated attributes of the first image. This process is done to ensure attribute annotation consistency across the images of the same speaker. Since the LRS dataset is used solely for research purposes, this usage is consistent with the Creative Commons Attribution 4.0 International License.

\begin{table}[t]
\centering
\resizebox{0.82\columnwidth}{!}{%
\begin{tabular}{llc}
\toprule[1.2pt]
\textbf{Attribute} & \textbf{Category} & \textbf{Percentage} \\
\midrule[1.0pt]
\multirow{2}{*}{Gender} & Male & 74.66\% \\
                        & Female & 25.34\% \\
\midrule[0.8pt]
\multirow{6}{*}{Race}   & Caucasian & 72.23\% \\
                        & Asian & 12.76\% \\
                        & Middle Eastern & 5.64\% \\
                        & African American & 4.72\% \\
                        & Latino Hispanic & 4.33\% \\
                        & Indian & 0.32\% \\
\bottomrule[1.2pt]
\end{tabular}
}
\caption{Dataset statistics according to gender and race attributes.}
\label{tab:distribution1}
\end{table}

\begin{table}[t]
\centering
\resizebox{0.80\columnwidth}{!}{%
\begin{tabular}{llc}
\toprule[1.2pt]
\textbf{Attribute} & \textbf{Category} & \textbf{Spk Sim} \\
\midrule[1.0pt]
\multirow{2}{*}{Gender} & Male & 0.7597 \\
                        & Female & 0.7503 \\
\midrule[0.8pt]
\multirow{6}{*}{Race}   & Caucasian & 0.7593 \\
                        & Asian & 0.7637 \\
                        & Middle Eastern & 0.6925 \\
                        & African American & 0.7165 \\
                        & Latino Hispanic & 0.7867 \\
                        & Indian & 0.7631 \\
\bottomrule[1.2pt]
\end{tabular}
}
\caption{Speaker similarity scores across gender and race.}
\label{tab:distribution2}
\end{table}

\section{Sub-attribute Category Evaluations}
\label{sec:attribute_classification}
To incorporate high-level attributes across both audio and visual modalities, we have introduced a bilateral enhancement strategy\footnote{The weight assigned to classification in the acoustic domain is empirically set to 0.3.}. To assess the robustness of our method across demographic subgroups, we analyze speaker similarity (SECS) scores by gender and race\footnote{The pretrained DeepFace model originally labels race attributes using terms such as "White" and "Black." For improved clarity and alignment with academic conventions, we adopt the terms "Caucasian" and "African American" in this paper.} using pseudo-attribute annotations derived from a pretrained \texttt{DeepFace} model on the LRS dataset. As summarized in Table~\ref{tab:distribution1}, the majority of the face images are classified as male (74.66\%) and predominantly Caucasian (72.23\%). Despite this imbalance, our model maintains consistent speaker similarity across subgroups, as shown in Table~\ref{tab:distribution2}. Specifically, the SECS scores for male (75.97) and female (75.03) speakers are closely aligned, which suggests that the model does not disproportionately favor the overrepresented gender.

\begin{figure}[t]
  \centering
  \includegraphics[height=4.7cm]{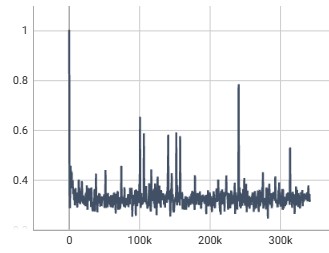}
  \caption{Visualization of collective modality attribute loss.}
  \label{fig:loss}
\end{figure}

A similar trend is observed across race categories; while Caucasian speakers dominate the dataset, the similarity scores for underrepresented groups such as Asian (76.37), Indian (76.31), and Latino Hispanic (78.67) are comparable or even superior. Notably, the performance for Middle Eastern (69.25) and African American (71.65) speakers, though slightly lower, still remains within an acceptable range, which indicates generalizability beyond the majority class. These findings are further supported by the progression of the attribute classification loss (Figure \ref{fig:loss}); its continuous decline and eventual plateau indicate stable convergence and the effectiveness of incorporating attribute supervision in improving face-to-voice alignment.

\begin{figure}[t]
    \centering
    \begin{minipage}{0.48\columnwidth}
        \centering
        \includegraphics[width=\linewidth, height=3.7cm]{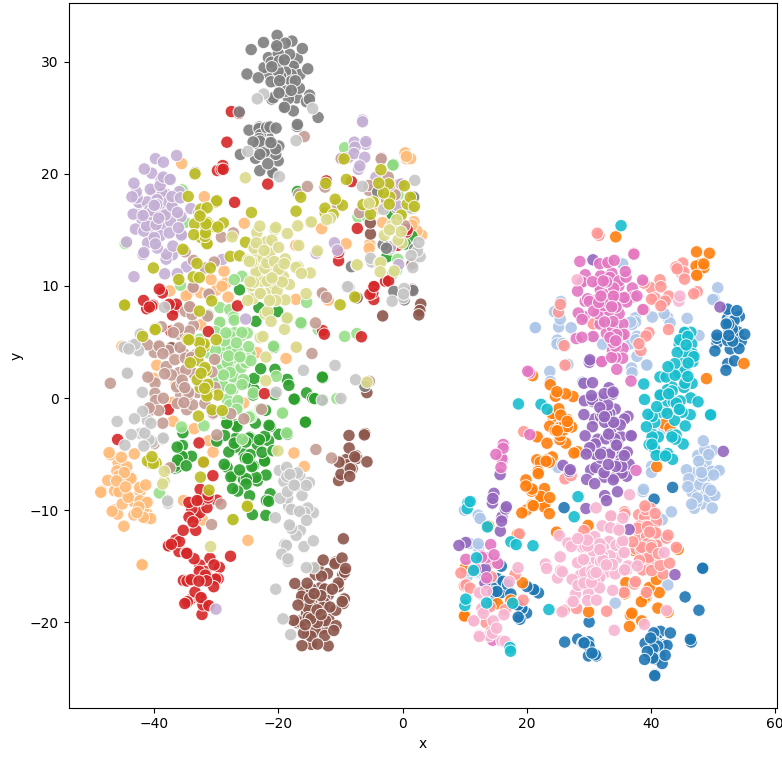}
        \caption*{(a) Pluster*}
    \end{minipage}
    \hfill
    \begin{minipage}{0.48\columnwidth}
        \centering
        \includegraphics[width=\linewidth, height=3.7cm]{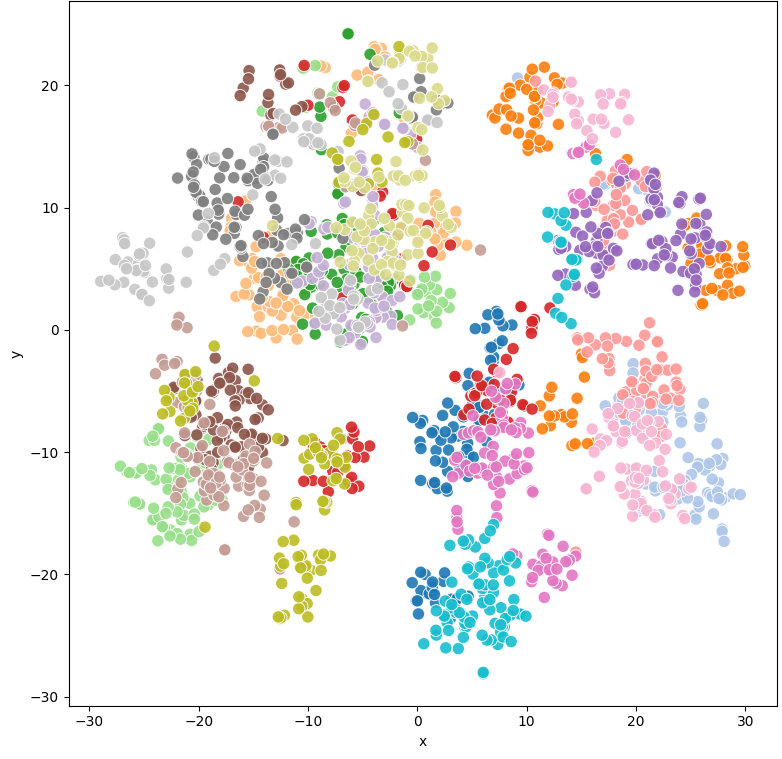}
        \caption*{(b) FaceTTS}
    \end{minipage}

    \vspace{0.5em}

    \begin{minipage}{0.48\columnwidth}
        \centering
        \includegraphics[width=\linewidth, height=3.7cm]{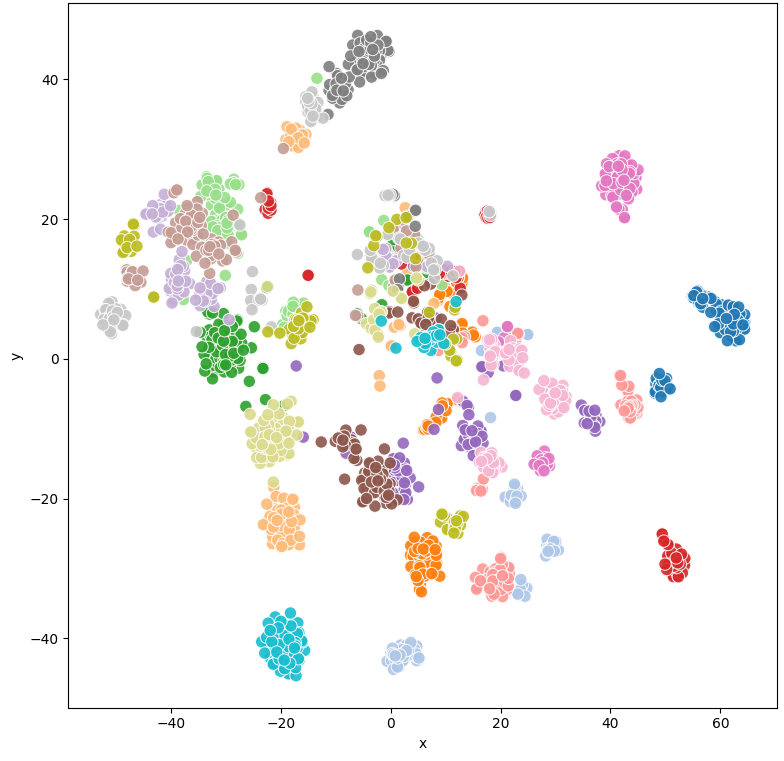}
        \caption*{(c) FVTTS}
    \end{minipage}
    \hfill
    \begin{minipage}{0.48\columnwidth}
        \centering
        \includegraphics[width=\linewidth, height=3.7cm]{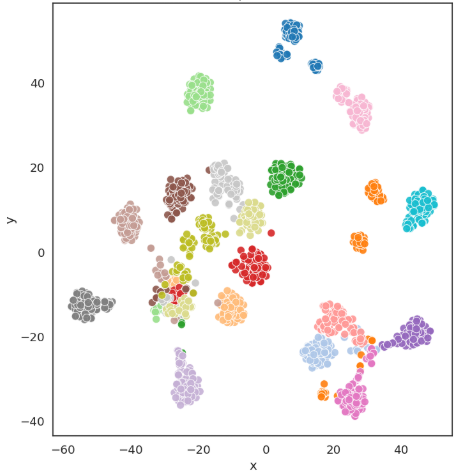}
        \caption*{(d) Ours}
    \end{minipage}

    \caption{t-SNE visualizations of speech embeddings generated using Resemblyzer for unseen speakers.}
    \label{fig:ablation-vis2}
\end{figure}

\section{Subjective Evaluations}
\label{sec:subjective}
To comprehensively assess the perceptual quality and speaker fidelity of the generated speech, each model was evaluated using 100 audio samples spanning 40 speakers, comprising of both seen and unseen identities. The evaluation process required approximately two hours per annotator. Among the twenty-one annotators, the gender distribution was seventeen male and four female. Prior to evaluation, all participants were informed that their demographic information and responses would be used solely for research purposes. To further substantiate our subjective assessments, we additionally provide objective visualizations of the unseen speakers in Figure~\ref{fig:ablation-vis2}.

\end{document}

%% file: introduction.tex
\section{Introduction}
The ability to communicate using one's own voice is an intrinsic and fundamental aspect of human identity, self-expression, and social interaction. However, a range of neurological and physiological conditions can severely impair speech production mechanisms. For instance, dysarthria is a speech disorder that results from various etiological factors, including cerebrovascular incidents (e.g., strokes) and degenerative neuromuscular disorders such as multiple sclerosis and Parkinson’s disease \cite{Cause2, Cause1}. The severity of dysarthria varies, but nevertheless manifests as slurred, unintelligible, and phonetically distorted speech. In more extreme cases such as total glossectomy or orofacial myofunctional disorders, individuals may completely lose the ability to generate speech. These impairments significantly hinder verbal communication, often leading to frustration, social isolation, and a reduced quality of life \cite{Life1}.

Towards this, text-to-speech (TTS) systems \cite{Tacotron2, Fastspeech2, vits} have been utilized as assistive technologies to convert typed text into artificial speech \cite{Life1, Life2}. However, while  they are effective for communication, they fail to preserve the speaker’s unique vocal identity. Due to this, personalized multi-speaker TTS methods \cite{cv2, multispeaker, cv3} could be a potential solution \cite{Life1} as they are able to imitate a target speaker's vocal characteristics for speech generation. Yet, achieving high-fidelity voice cloning typically requires extensive multi-speaker training or few-shot adaptation with a large amount of speech recordings. Thus, while zero-shot multi-speaker synthesis can ideally enable speaker adaptation, it remains infeasible for those without any accessible prior recordings, such as individuals with complete vocal muscle paralysis. This limitation highlights the need for alternative biometric modalities to infer speaker identity and enable personalized voice synthesis that is independent of speech recordings.

Both early and recent psychological research suggests the existence of a strong correlation between facial identity with vocal characteristics. Specifically, studies such as \citet{psychological2, psychological3, psychological1} have indicated that humans can infer aspects of a person’s voice, such as pitch and timbre, based solely on their facial features. As a result, this insight has driven investigations into face-to-voice (FTV) synthesis, where a facial image serves as a reference for generating a corresponding voice, and then producing speech in that inferred voice.

Early methods of FTV synthesis relied on statistical techniques that mapped facial images onto facial landmarks, which were then transformed into eigenvoice representations \cite{eigenvoice1}. These approaches, however, were limited by the need for manual alignment between facial and vocal features. More recent work has adopted multi-stage training pipelines \cite{pretrained1, 3Dface, multi-stage2}, wherein a face encoder is first trained to produce embeddings that align with outputs from a separately trained audio-based speech encoder. Afterwards, during inference, the face encoder substitutes the speech encoder. While this strategy improves performance, it introduces training complexity and inefficiency. Moreover, because the facial representations are specifically trained to align with audio features, they fail to directly capture and utilize fine-grained local facial features such as gender or other demographic attributes that are essential in producing identity-consistent speech. Although \citet{3Dface} attempts to explicitly incorporate such facial attributes, their method requires a combination of demographic metadata, 2D facial features, textures (e.g., skin, muscle), and 3D cranial structures. Not only does this introduce substantial complexity in terms of data acquisition and input modalities, but also necessitates a three-stage learning process to integrate these sources of information, which ultimately results in a complex and resource-intensive training pipeline.

To address the limitations of prior work, specifically the reliance of inference-time-only visual encoders, underutilization of fine-grained facial features, and dependence on multi-stage training, we present a novel end-to-end FTV synthesis framework, which eschews external models and instead focuses on effective facial feature learning. Our method starts with a progressive facial encoder that decomposes and hierarchically aggregates local facial regions to form a rich visual identity embedding. Additionally, to ensure alignment of high-level semantic attributes such as gender and ethnicity, we introduce a bilateral attribute enhancement mechanism, which is applied to both the facial representation and to the synthesized audio output. A multi-view data augmentation strategy is further adopted where each speech sample is paired with a diverse set of facial images from the same speaker, which are captured under varying poses, lighting conditions, and facial movements. Through both subjective and objective evaluations, we demonstrate that our method generates speech with higher speaker fidelity and stronger FTV associations.

%% file: relatedwork.tex
\section{Related Work}
\subsection{Stylistic Speech Generation}
Advancements in TTS technology have greatly enhanced the naturalness of synthesized speech, thereby expanding its applications to areas such as voice imitation in the form of multi-speaker TTS. Methodologies of this domain incorporate a speaker encoder that extracts a speaker representation from a reference audio sample, and then conditions it into the backbone TTS model. Existing methodologies typically follow either a few-shot adaptation strategy, where a pre-trained multi-speaker TTS model is fine-tuned with multiple target speaker samples \cite{FS-1, FS-2, FS-3}, or a zero-shot adaptation approach, which generates an embedding from a single target speaker sample to condition the model directly. Regardless of the adaptation objective, research in multi-speaker TTS has either focused on improving speaker encoders \cite{ZS-1, cv3} or refining speaker conditioning techniques by evolving from simple concatenation to more adaptive methods \cite{SALN, SNAC, sc-cnn}. While FTV synthesis is a form of multi-speaker TTS, it takes a fundamentally different approach. Instead of relying on reference audio at training or inference time, it leverages a facial image to infer a speaker's vocal characteristics. As such, this task is inherently more challenging, as it requires learning to associate visual identity cues such as gender, with corresponding acoustic traits.

\begin{figure*}[t]
  \centering
  \includegraphics[width=\textwidth, height=10.5cm]{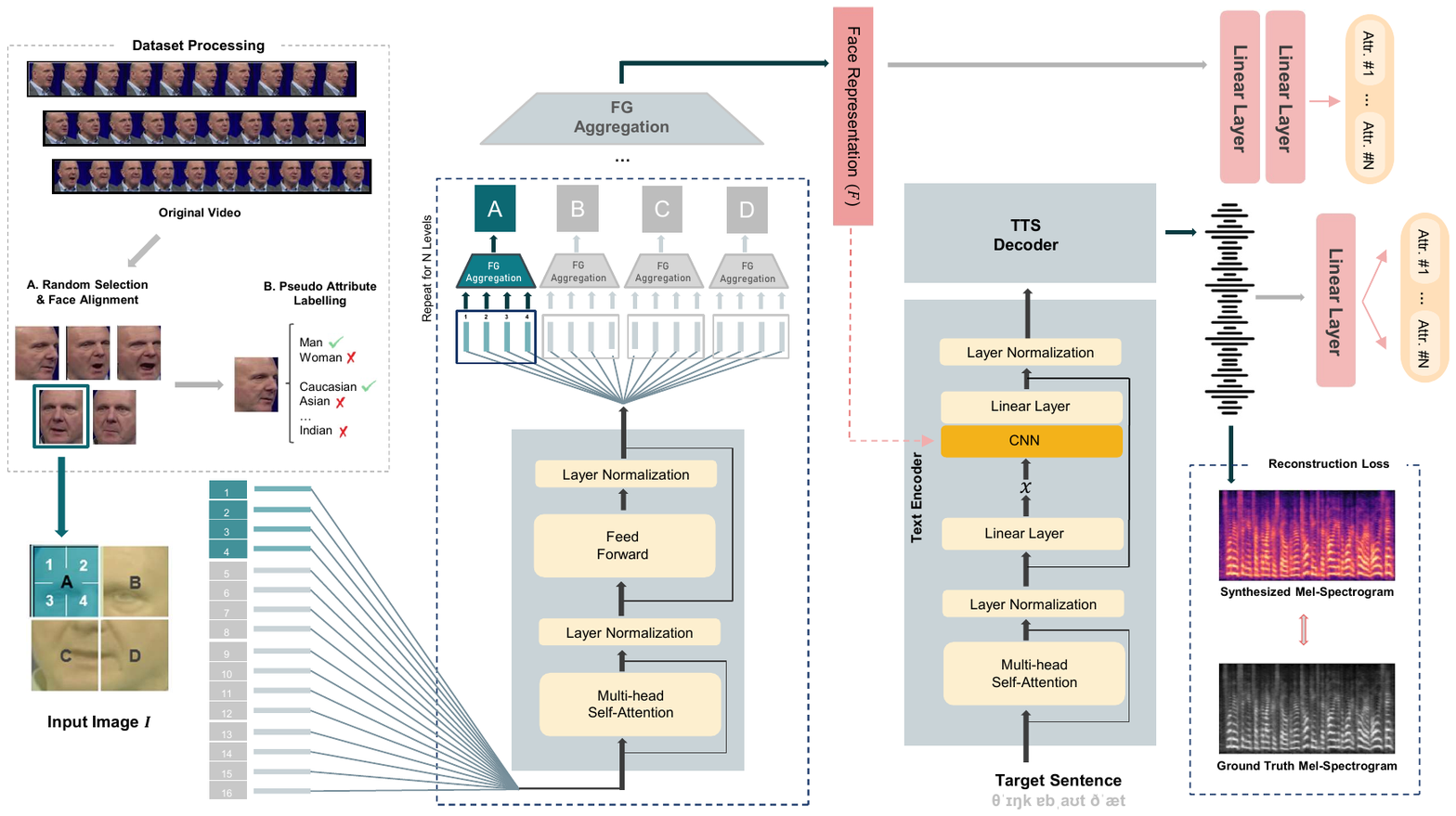}
  \caption{Diagram illustrating the proposed model. On a high level, an input image is segmented into 16 smallest-size patches. Four sets of adjacent patches are then combined to create four larger areas. These four areas are subsequently aggregated into a final representation $F$, which ultimately integrates information from all 16 original patches. Moreover, attribute enhancement is further conducted on both the facial and acoustic domains. Note that the architecture following the text encoder output is identical to the original VITS \cite{vits} model.}
  \label{fig:arch}
\end{figure*}

\subsection{Multi-Modal Speech Synthesis}
The innate human ability to associate voices with faces \cite{Neuro1}, along with evidence of overlapping neural-cognitive pathways for processing these modalities \cite{Neuro2}, has fueled extensive research in multi-modal visual-acoustic learning. A prominent direction in this field is voice-to-face generation, where models synthesize plausible facial features from speech, effectively constructing a visual identity from audio \cite{TalkingFace1, TalkingFace2, Review-paper2}. In contrast, face-to-voice generation has been studied in applications like biometric security \cite{biometrics}, lip-dubbing \cite{lipsync, lipsync2, lipsync3, lipsync4, lipsync5} and voice synthesis \cite{3Dface, multi-stage3, multi-stage2}. Previous approaches in voice synthesis often rely on multi-stage training pipelines by aligning face and audio embeddings, and then only leveraging face encoders during inference \cite{pretrained1, multi-stage2}. To enhance identity preservation across modalities, various loss functions such as tri-item \cite{pretrained2} and disentanglement losses have also been formulated \cite{faceLoss1}. In this work, we focus on face-to-voice synthesis while addressing the limitations of previous methods by eliminating multi-stage training. Towards this objective, we extract effective facial embeddings enriched with identity-relevant attributes like gender and ethnicity for higher speaker fidelity. 

%% file: method.tex
\section{Methodology}
\subsection{Problem Setup}
In a conventional multi-speaker TTS framework, the training dataset is defined as $D = \{(X_n, Y_n)\}_{n=1}^N$, where $X_n$ and $Y_n$ represent the input text and corresponding speech waveform, respectively. The objective is to generate an acoustic output $\hat{Y}_n$ that reflects both the linguistic content of $X_n$ and the vocal identity of the target speaker. To enable speaker-specific synthesis, an auxiliary speaker encoder is employed to extract a speaker embedding from the reference audio $Y_n$.

Extending this framework to the FTV synthesis task, we aim to generate speech that aligns with the input text and conveys the vocal characteristics of the individual, which is given as a facial image. Accordingly, the training dataset is reformulated as $D = \{(X_n, Y_n, Z_n)\}_{n=1}^N$, where $Z_n$ denotes the facial image that is extracted from a video. A facial representation $F$ is then extracted from $Z_n$ and conditioned into the end-to-end acoustic model of the VITS architecture~\cite{vits}. Following the adaptive conditioning approach of~\citet{sc-cnn}, $F$ is projected through a linear layer to obtain a gain $g$, a normalized direction vector $\frac{d}{\lVert d \rVert}$, and a bias term $b$. These parameters are used in a 1-D convolutional layer to modulate the phonemic hidden representations from the VITS text encoder $x$ as $(g \times \frac{d}{\lVert d \rVert}) \times x + b.$ This fusion mechanism integrates facial identity cues into the linguistic representation, which guides the model to synthesize speech that is both textually accurate and identity-consistent. Since representation conditioning strategies are a separate and active area of research \cite{sc-cnn, SALN, SNAC}, we do not elaborate on this in further detail. Interested readers are referred to \citet{sc-cnn}. In the following subsections, we instead focus on the learning of facial representation $F$. The full architecture is illustrated in Figure \ref{fig:arch}.

\subsection{Progressive Feature Extraction}

Although the primary objective of FTV is to convert facial images into speech, previous literature have predominantly emphasized audio signal decomposition over image processing. This is exemplified by the learning of latent acoustic characteristics from reference audio, which is subsequently used as a target for training the face encoder \cite{pretrained1, pretrained2, multi-stage2}. Such strategies emphasize global representations and often overlook the rich spatial hierarchies embedded within facial images. Towards this, we incorporate a hierarchical non-pretrained visual encoder originally utilized for image generation \cite{nest} that is trained directly from scratch, and results in more precise and robust mapping from facial geometry to vocal identity.

Formally, given a facial input image \( Z \in \mathbb{R}^{3 \times H \times W} \), the model first decomposes \( Z \) into non-overlapping patches of resolution \( P \times P \). Each patch \( z_{i,j} \in \mathbb{R}^{3 \times P \times P} \) is flattened and linearly projected into a fixed-dimensional embedding space, which results in a set of patch tokens:

\scalebox{0.87}{\parbox{\linewidth}{%
    \begin{align}
    Z = \{ z_{i,j} \mid i \in [1, H/P], \, j \in [1, W/P] \}
    \end{align}
}}

Each patch token is then passed through canonical Transformers~\cite{transformers} that is composed of alternating multi-head self-attention (MHSA), feed-forward networks (FFNs), and Layer Normalization (LN):

\scalebox{0.87}{\parbox{\linewidth}{%
    \begin{align}
    \hat{z} = \text{TransformerBlock}(z)
    \end{align}
}}

Moreover, it is important to conduct aggregation between neighboring granularities $\hat{z}$ in order to form a comprehensive visual representation and to enhance locality. As such, the first facial granularity aggregation (FGA) iteration combines the representations of the four smallest neighboring $P \times P$ patches using a 3 $\times$ 3 convolutional layer, followed by layer normalization, and then 3 $\times$ 3 max pooling. This process expands the model's receptive area from a single local facial patch to a set of four.

\scalebox{0.87}{\parbox{\linewidth}{%
    \begin{align}
   \tilde{z} = \text{MaxPool}(\text{LN}(\text{Conv}_{3 \times 3}([\hat{z}_{1}, \cdots, \hat{z}_{n})))
    \end{align}
}}

This Transformer–aggregation cycle is applied iteratively across hierarchical levels, where each aggregation stage reduces the number of spatial patches by a factor of four, progressively expanding the receptive field. The process continues until a single spatial token remains, thus capturing the global summary of the entire facial image. The final output is a 128-dimensional embedding that serves as the facial identity vector $F$.

\subsection{Bilateral Attribute-Based Enhancement}

While the hierarchical visual encoder described in the previous section effectively captures spatially localized geometric features, it does not account for high-level semantic attributes that humans often leverage when inferring vocal traits from facial appearance. Specifically, psychological studies suggest that demographic cues such as gender and ethnicity influence perceived voice characteristics \cite{attribute-Nationality1, attribute-Nationality2}. Thus, to complement the recursive visual learning of facial structure, we introduce an auxiliary attribute-based learning objective that explicitly aligns and enhances high-level facial semantics such as gender and ethnicity into the training process. 

Towards this, we employ a pseudo-labeling approach using the pretrained DeepFace model\footnote{\url{https://github.com/serengil/deepface}}  \cite{deepface}, which automatically annotates each facial image with predicted demographic categories pertaining to gender and race. Attribute prediction is then conducted across both visual and auditory modalities to enhance cross-modal consistency. On the visual side, the facial representation \( F \) is passed through two fully connected layers to predict an attribute, which is then compared against the pseudo-labeled ground truth via a standard cross-entropy loss. Facial embedding \( F \) is subsequently integrated into the backbone TTS system using adaptive conditioning techniques from \citet{sc-cnn}. On the auditory side, the synthesized waveform of variable lengths is first temporally linearly resampled into a fixed-size vector representation. A separate linear layer then predicts the same set of attributes, which enables semantic supervision in the audio domain. This bidirectional supervision scheme promotes alignment of high-level identity cues across modalities. The overall attribute prediction loss is

\begin{table}[t]
\centering
\resizebox{7.5cm}{!}{
\begin{tabular}{ccc}
\toprule[1.2pt]
\textbf{Evaluation Metric} & \textbf{Modality} & \textbf{Comparison Set} \\ \midrule[1.0pt]
\textbf{MOS} & Audio $\Leftrightarrow$ Audio & 1 : 1\\ 
\textbf{ABX} & Audio $\Leftrightarrow$ Audio & 1 : 4\\ 
\textbf{F2V Alignment} & Image $\Rightarrow$ Audio & 1 : 4\\ 
\textbf{Age Alignment} & Audio & \textemdash \\ 
\textbf{Gender Classification} & Audio &  \textemdash \\ \bottomrule[1.2pt]
\end{tabular}}
\caption{Overview of subjective evaluation methodologies. \textit{Modality}: MOS, ABX, and Face-to-Voice (F2V) Alignment evaluations involve comparing the synthesized audio output with either the ground truth audio or a corresponding face image. \textit{Comparison Set}: Indicates the number of synthesized audio samples that are evaluated against each other to determine the one most closely aligned with the ground truth.}
\label{tab:evaluation-summary}
\end{table}

\scalebox{0.87}{\parbox{\linewidth}{%
    \begin{align}
   \mathcal{L}_{\text{attr}} = \mathcal{L}_{\text{face}} + \alpha \cdot \mathcal{L}_{\text{audio}}
    \end{align}
}}
where \( \mathcal{L}_{\text{face}} \) and \( \mathcal{L}_{\text{voice}} \) denote the cross-entropy losses for the attribute predictions from the facial and voice modalities, respectively, and are combined through weighted summation. These losses are linearly combined with the original reconstruction loss calculated between ground truth and synthetic mel-spectrograms, along with the KL-divergence, adversarial, and duration losses utilized in VITS \cite{vits} for training.

Moreover, to enhance robust learning, multi-view data augmentation of the training data is further conducted. This is because previous methodologies have only relied on datasets featuring static relationships \cite{pretrained1, adapt1, adapt2, multi-stage2}, wherein each speaker is represented by a single face image paired with corresponding ground truth audio for FTV model training. Recognizing the need for vocal consistency despite variations in head orientation and facial expression, we employ a pretrained s3fd \cite{s3fd} face alignment model to automatically extract face-centered frames from each video in the LRS \cite{lrs3} training dataset. From these frames, five images are randomly selected and paired with the identical corresponding audio to train the previously detailed model. An overview of this process is provided in the upper left side of Figure \ref{fig:arch}.\footnote{Further processing details are provided in Appendix \ref{sec:dataset_processing}.}

%% file: experiments.tex
\section{Experimental Settings}
We leverage the trainval subset of the LRS open-source dataset \cite{lrs3} for training, and conduct dataset multi-view augmentation and pseudo-labelling\footnote{We do not use age attributes due to significant bias, as the data is heavily concentrated in the 30–40 age range.} as detailed in Sections 3.2 and 3.3, respectively. Out of a total of 159,511 utterances, the training-validation is split using a ratio of 9:1. Audios are resampled to 16,000 Hz, and we use a filter and window length of 1024, and hop size of 256 for mel-spectrogram processing. For fair comparisons, all experiments adhere to a batch size of 20, and 350,000 training iterations are conducted for approximately two days using a single A6000 GPU. The total number of parameters is approximately 43.6 million.

\begin{table*}[t]
\centering
\resizebox{0.92\linewidth}{!}{%
\begin{tabular}{lccccccc}
\toprule[1.2pt]
\multirow{2}{*}{\textbf{Model} \vspace{-0.5em}} & \multicolumn{3}{c}{\textbf{Subjective Metrics}} & \multicolumn{4}{c}{\textbf{Objective Metrics}} \\
\cmidrule(r){2-4} \cmidrule(l){5-8}
& \textbf{MOS ($\uparrow$)} & \textbf{ABX ($\uparrow$)} & \textbf{F2V ($\uparrow$)} & \textbf{Seen SECS ($\uparrow$)} & \textbf{Unseen SECS ($\uparrow$)} & \textbf{CER ($\downarrow$)} & \textbf{UTMOS ($\uparrow$)} \\ \midrule[1.0pt]
Pluster*  & 3.25$\pm$0.08  & 20.83\% & 16.55\% & 67.04    & 64.81   & 0.2381 & 3.0470         \\ 
FaceTTS  & 2.98$\pm$0.09  & 28.45\%  & 25.71\% & 60.11      & 56.54   & \textbf{0.1070}     & 2.1268    \\ 
FVTTS   & 3.20$\pm$0.08  & 19.64\%  & 27.38\% & 62.50      & 59.49   &  0.2743    & 2.2620   \\ 
\begin{tabular}[c]{@{}l@{}}Ours \end{tabular} & \textbf{3.51$\pm$0.09} & \textbf{31.07\%} & \textbf{30.36\%} & \textbf{79.96}  &\textbf{71.39}   & 0.1302      & \textbf{3.2218}  \\ \bottomrule[1.2pt]
\end{tabular}%
}
\caption{Objective and subjective metric evaluation results. MOS scores are calculated with 95\% confidence. F2V denotes the face-to-voice alignment metric.}
\label{tab:main-results}
\end{table*}

\subsection{Evaluation Protocol}
For validation, we juxtapose our method with three baseline systems, and conduct subjective and objective assessments. The first baseline follows the two-stage training method proposed by \citet{adapt2}, which jointly trains a Global Style Token \cite{GST} speaker encoder with the TTS backbone. Concurrently, a pretrained visual encoder is finetuned to align with the learned speaker embeddings\footnote{Although the original implementation is based on Tacotron, we re-implement the same methodology using the VITS architecture, and refer to this model as Pluster*.}. The second model is a diffusion-based framework  (\citet{adapt1}, FaceTTS) that incorporates an auxiliary speaker alignment loss that encourages the visual and acoustic embeddings of the same identity to be close in a shared latent space. Lastly, FVTTS \cite{FVTTS} combines both global and local facial representations; a pretrained facial recognition network FaceNet \cite{facenet} is used to to extract global embeddings, while an additional face encoder is utilized to extract local features from the same input image.

\begin{table}[t]
\centering
\resizebox{0.8\linewidth}{!}{%
\begin{tabular}{lccc}
\toprule[1.2pt]
         \textbf{Model}& \textbf{Younger} & \textbf{Identical} & \textbf{Older} \\ \midrule[1.0pt]
Pluster* & 0.225   & 0.538    & 0.237 \\ 
FaceTTS  & 0.156   & 0.520    & 0.324 \\ 
FVTTS   & 0.255   & 0.508     & 0.237 \\ 
Ours    & 0.164   & 0.649     & 0.187 \\ \bottomrule[1.2pt]
\end{tabular}%
}
\caption{Age alignment accuracy expressed as a percentage. The inferred age of the voice in the synthesized speech is classified as younger, identical, or older relative to the ground truth audio.}
  \label{fig:age}
\end{table}

To comprehensively assess the perceptual quality and speaker fidelity of the generated speech, we conduct a range of subjective evaluations via Amazon Mechanical Turk\footnote{Participants were compensated according to the hourly wage in the authors' country of residence.} with twenty-one participants. The first subjective evaluation task involved participants assessing vocal similarity between synthesized and ground truth audio using a 5-point Likert scale (Mean Opinion Scores (MOS)). In the second ABX task, participants selected the audio sample from a set of four (each generated by a different model) that best matched the target speaker's ground truth audio. 

While traditional evaluations in this domain typically rely on just MOS and ABX tests, we expand our protocol (Table \ref{tab:evaluation-summary}) to include face-to-voice (F2V) and age alignment, and gender classification tasks to better analyze speaker identity preservation. Specifically, the third task of F2V alignment required participants to choose one audio sample out the four audios generated from each of the three baselines and the proposed model, that best matched a given target speaker image. For the age classification task, participants categorized each synthesized audio as having a younger, older, or similar vocal age compared to the ground truth audio. Finally, in the gender classification task, each synthesized audio sample were indicated as either having a male or female voice.

In addition, to support the validity of the subjective evaluation scores, we incorporate a set of automatic objective metrics. Speaker similarity is quantified via cosine similarity between the ground truth target speaker and the synthesized audio (SECS), using the Resemblyzer\footnote{https://github.com/resemble-ai/Resemblyzer} package. To substantiate the naturalness of the synthesized speech, we employ UTMOS\footnote{https://github.com/sarulab-speech/UTMOS22} \cite{UTMOS}, which is a pre-trained model that is designed to predict MOS scores for synthesized speech. Lastly, given that FTV synthesis is fundamentally a speech generation task, maintaining clear and accurate pronunciation is essential. To objectively evaluate pronunciation quality, we transcribe the synthesized speech using a pretrained Wav2Vec-base-960h \cite{w2v} model, and then compute the Character Error Rate (CER) with the jiwer\footnote{https://pypi.org/project/jiwer/} library.

\begin{figure}[t]
  \centering
  \includegraphics[width=0.9\linewidth, height=5cm]{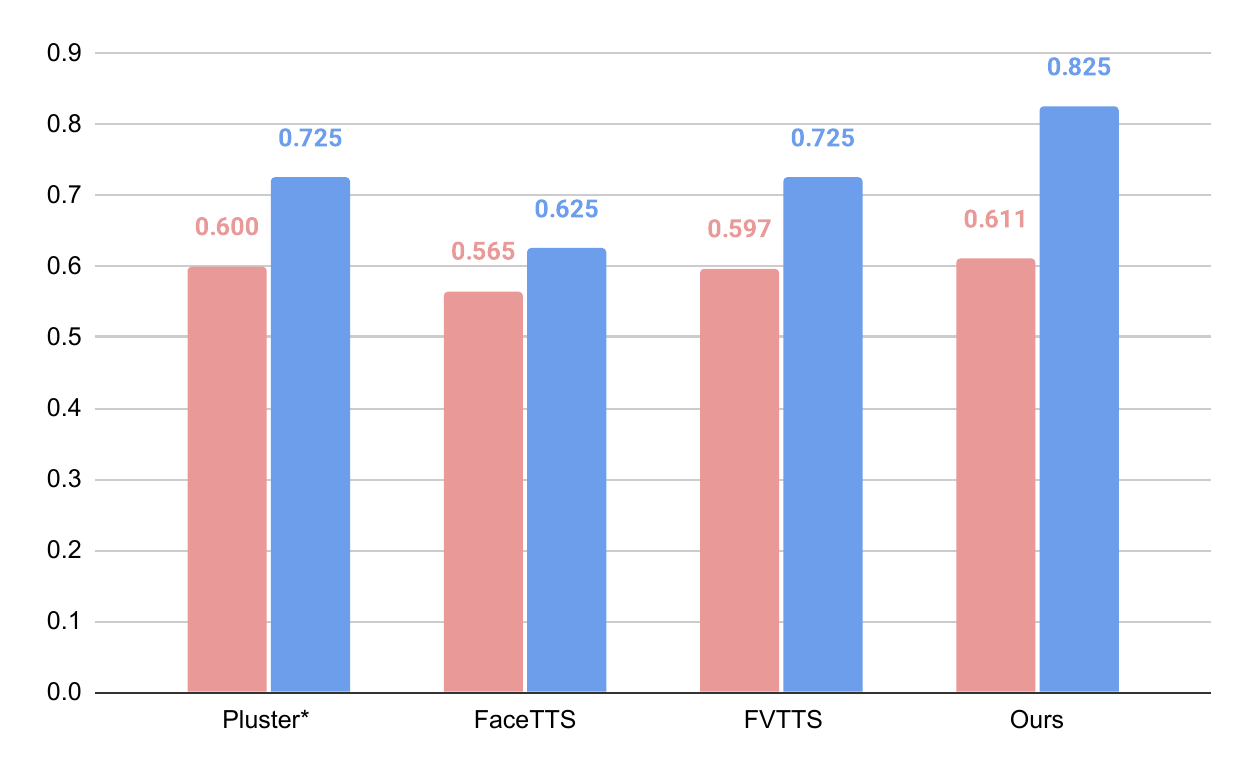}
  \caption{Gender classification agreement (pink) and accuracy (blue) scores in percentage. \textit{Agreement} is quantified by the number of annotator votes assigning the synthesized voice to a specific gender (male or female). \textit{Accuracy} is determined based on the majority vote outcome relative to the ground truth gender.}
  \label{fig:gender}
\end{figure}

%% file: results.tex
\section{Results and Analysis}

\begin{figure}[t]
    \centering
    \begin{minipage}{0.48\columnwidth}
        \centering
        \includegraphics[width=\linewidth, height=3.7cm]{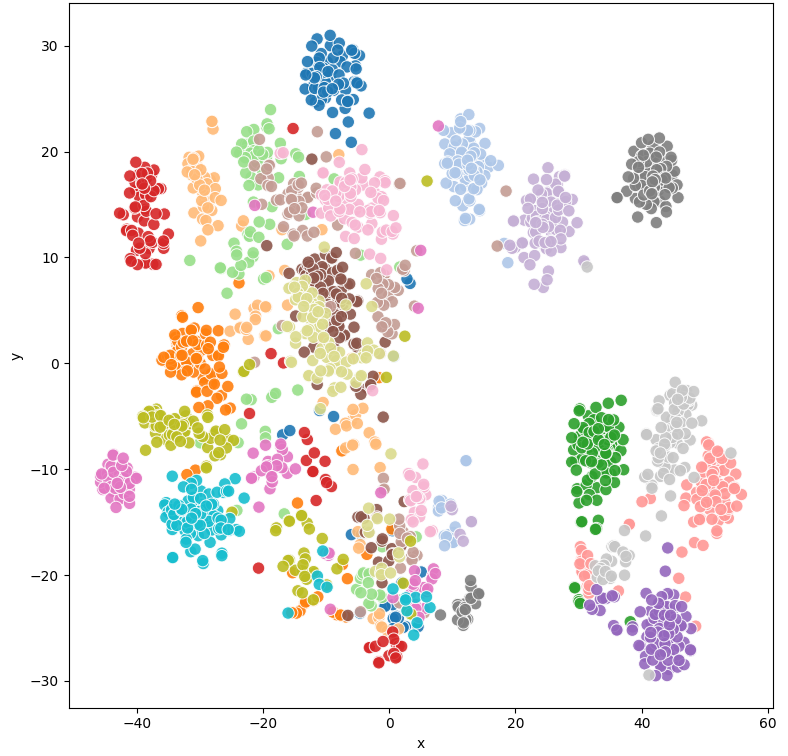}
        \caption*{(a) Pluster*}
    \end{minipage}
    \hfill
    \begin{minipage}{0.48\columnwidth}
        \centering
        \includegraphics[width=\linewidth, height=3.7cm]{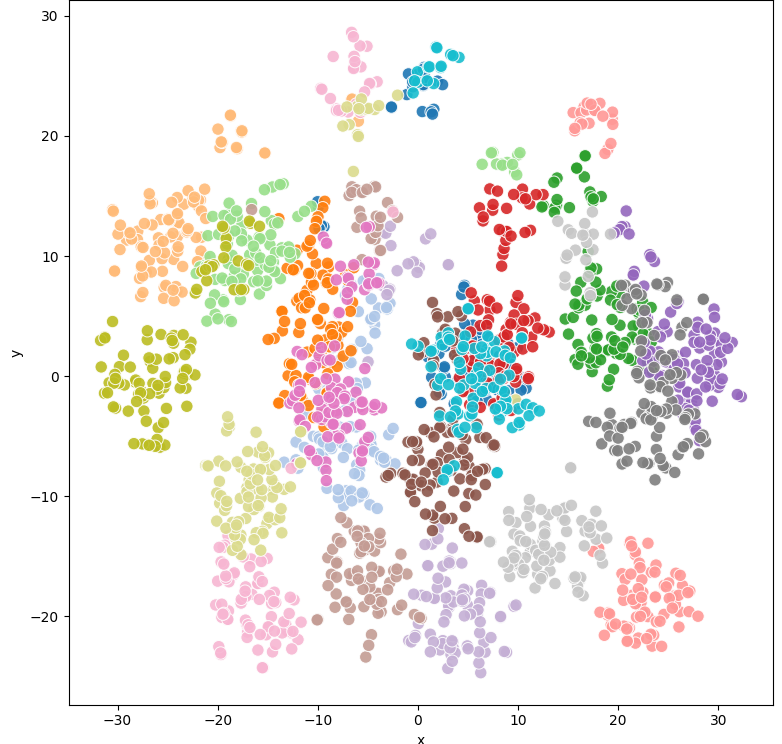}
        \caption*{(b) FaceTTS}
    \end{minipage}

    \vspace{0.5em}

    \begin{minipage}{0.48\columnwidth}
        \centering
        \includegraphics[width=\linewidth, height=3.7cm]{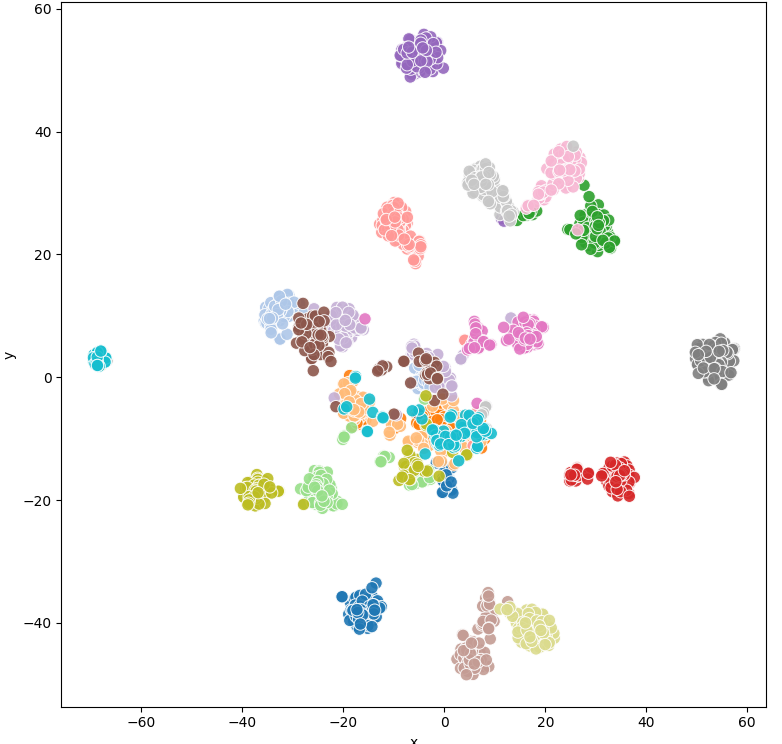}
        \caption*{(c) FVTTS}
    \end{minipage}
    \hfill
    \begin{minipage}{0.48\columnwidth}
        \centering
        \includegraphics[width=\linewidth, height=3.7cm]{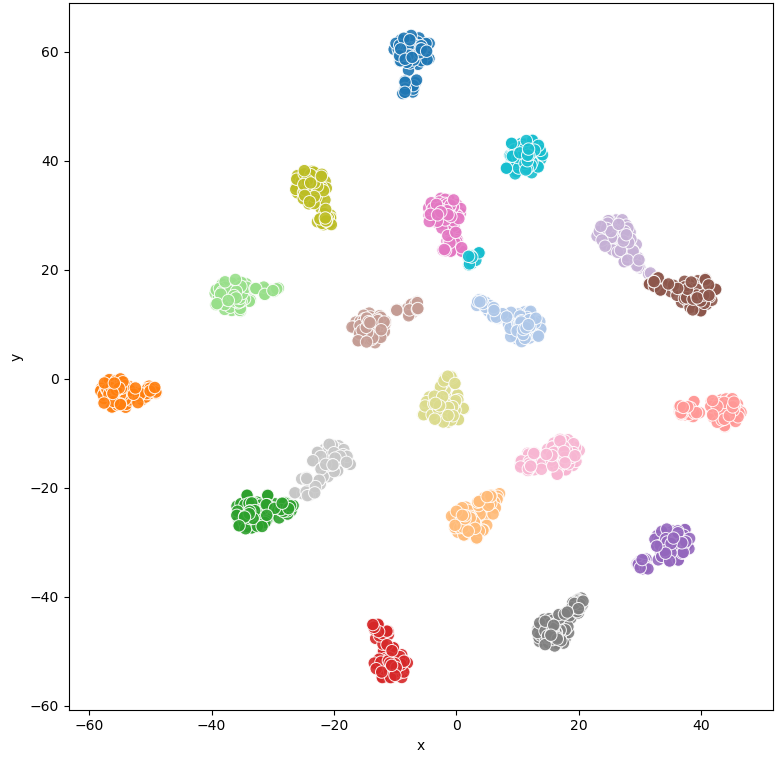}
        \caption*{(d) Ours}
    \end{minipage}

    \caption{t-SNE visualizations of speech embeddings generated using Resemblyzer for seen speakers. Each color denotes a distinct speaker identity. Compact and well-separated clusters indicate higher speaker consistency and better identity preservation. Visualizations of unseen speakers (Figure \ref{fig:ablation-vis2}) are included in Appendix \ref{sec:subjective}.}
    \label{fig:ablation}
\end{figure}

\subsection{Feature Extraction and Speaker Fidelity}
The results in Table \ref{tab:main-results} clearly demonstrate the superior performance of our proposed model across both subjective and objective evaluation metrics. In terms of MOS, our model achieves a score of 3.51 with a notable 0.26 difference than the closest performing baseline (\citet{adapt2}, 3.25), which indicates significantly higher speaker fidelity in 1:1 comparisons with ground truth reference recordings. This is further reinforced by ABX preference scores where our model achieves 31.07\%, which indicates stronger perceptual similarity to the reference voices in 1:4 comparisons. Most critically, in the F2V alignment task, which evaluates the appropriateness of the inferred voice given only facial inputs, our model records the highest alignment score at 30.36\%. 

To better understand this perceptual alignment, we further examine age consistency via subjective preference scores across three categories: Younger, Identical, and Older (Table~\ref{fig:age}). Our model yields the highest preference for the Identical category, achieving 0.649, which is +0.111 higher than the best-performing baseline (\citet{adapt2}, 0.538). In contrast, baseline models exhibit a greater tendency to generate voices perceived as either too young or too old relative to the target, which suggests unstable synthesis and reduced identity coherence. These results underscore our model’s ability to generate voices that are more faithfully aligned with the target speaker’s perceived age. Additionally, in terms of gender appropriateness (Figure~\ref{fig:gender}), our model exhibits a 10\% improvement in classification accuracy and 2.4-point gain in agreement scores compared to the best-performing baseline.

\begin{figure}[t]
  \centering
  \includegraphics[width=0.9\linewidth, height=4.3cm]{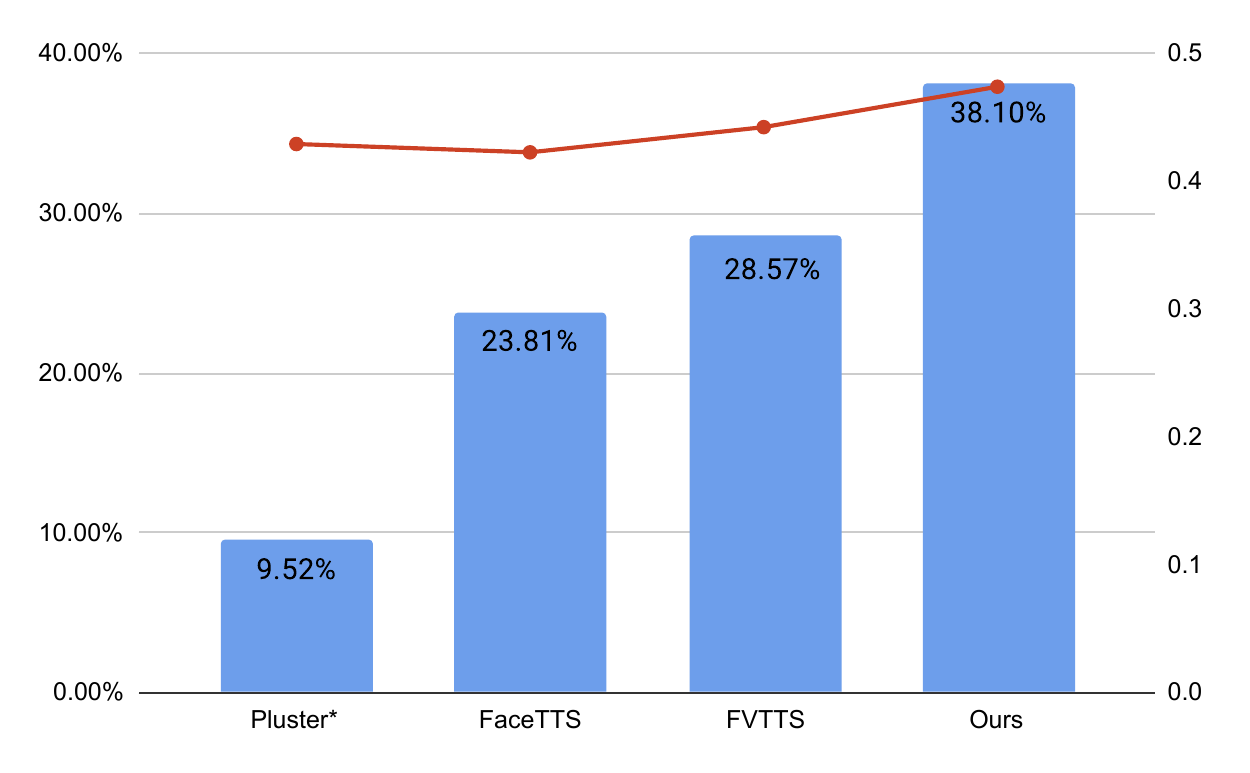}
  \caption{ABX (blue) and SECS (red) scores for the out-of-domain GRID dataset.}
  \label{fig:grid}
\end{figure}

These subjective findings are strongly corroborated by objective evaluations in Table~\ref{tab:main-results}. In SECS, which measures speaker embedding similarity, our model achieves the highest scores for both seen (79.96) and unseen speakers (71.39), with respective margins of 12.92 and 6.58 over the closest baseline model. Although \citet{adapt1} achieves a lower CER of 0.1070, the difference with our model is minimal at 0.0232. Furthermore, our model performs significantly better than \citet{adapt2} and \citet{FVTTS}, with CER differences of 0.1079 and 0.1441, respectively. In addition, in terms of naturalness, as measured with UTMOS, our model again leads with 3.2218, which indicates that our synthesized speech is perceived as more natural and human-like. These results are further supported by t-SNE \cite{TSNE} visualizations of speaker embeddings across different speakers, where tighter clusters of the same speaker demonstrate speaker consistency (Figure \ref{fig:ablation}).

\begin{table}[t]
\centering
\resizebox{0.95\columnwidth}{!}{%
\begin{tabular}{ccccc}
\toprule[1.2pt]
&\textbf{Attribute} & \textbf{Seen SECS ($\uparrow$)} & \textbf{Unseen SECS ($\uparrow$)} & \textbf{Overall} \\ \midrule[1.0pt]               
\multirow{3}{*}{\textbf{Image}}&Race                & \multicolumn{1}{c}{77.34} & \multicolumn{1}{c}{69.37} & \multicolumn{1}{c}{73.36} \\
&Gender              & \multicolumn{1}{c}{74.59} & \multicolumn{1}{c}{68.72} & \multicolumn{1}{c}{71.66} \\
&Race + Gender       & \multicolumn{1}{c}{\textbf{77.76}} & \multicolumn{1}{c}{\textbf{70.27}} & \multicolumn{1}{c}{\textbf{74.63}} \\ 
\midrule[0.7pt]
\multirow{3}{*}{\textbf{Audio}}&Race                & \multicolumn{1}{c}{75.44} & \multicolumn{1}{c}{67.57} & \multicolumn{1}{c}{71.51} \\ 
&Gender              & \multicolumn{1}{c}{75.28} & \multicolumn{1}{c}{68.48} & \multicolumn{1}{c}{71.88} \\
&Race + Gender       & \multicolumn{1}{c}{\textbf{75.84}} & \multicolumn{1}{c}{\textbf{69.52}} & \multicolumn{1}{c}{\textbf{72.48}} \\ 
\midrule[0.7pt]
\textbf{Image}&Race                & \multicolumn{1}{c}{76.14} & \multicolumn{1}{c}{68.34} & \multicolumn{1}{c}{72.24} \\ 
\textbf{+}&Gender              & \multicolumn{1}{c}{75.02} & \multicolumn{1}{c}{69.83} & \multicolumn{1}{c}{72.43} \\
\textbf{Audio}&Race + Gender       & \multicolumn{1}{c}{\textbf{79.96}} & \multicolumn{1}{c}{\textbf{71.39}} & \multicolumn{1}{c}{\textbf{75.68}} \\ \bottomrule[1.2pt]

\end{tabular}%
}
\caption{SECS-based speaker fidelity scores for both seen and unseen speakers under varying configurations of attribute-based supervision. Enhancement applied exclusively to the facial representation space is denoted as \texttt{Image}, while supervision constrained to the acoustic modality is indicated by \texttt{Audio}. The \texttt{Image+Audio} setting corresponds to joint attribute enhancement across both visual and auditory domains.}
\label{tab:loss1 ablation}
\end{table}

In order to evaluate generalization to out-of-domain settings, we conducted additional tests using the GRID dataset \cite{GRID}. As shown in Figure~\ref{fig:grid}, our model maintains the highest ABX preference score, with a +10\% margin over the second-best model. Additionally, we achieve the top SECS score at 47.38, representing a +3.16 improvement over the next best-performing system, further validating robustness.

\subsection{Attribute-based Enhancement}
The contribution of attribute-informed supervision, which is implemented via auxiliary prediction objectives for race and/or gender, is quantitatively substantiated by the SECS scores reported in Table~\ref{tab:loss1 ablation}. When applied exclusively to the facial representation space, unimodal conditioning on either race or gender yields modest improvements in speaker identity preservation, with overall SECS scores of 73.36 and 71.66, respectively. Their joint integration, however, results in an elevated score of 74.63, which is indicative of the complementary nature of multi-attribute cues in enhancing speaker fidelity. In contrast, restricting attribute-based enhancement to the acoustic modality results in a relatively small improvement, which suggests that facial embeddings provide a more robust supervisory signal for identity conditioning. Nonetheless, consistent with the visual-only configuration, combining race and gender supervision within the audio domain still outperforms single-attribute variants for both seen and unseen speaker subsets. 

The most pronounced gains emerge when attribute supervision is concurrently applied across both modalities. The \texttt{Race+Gender} configuration in this dual-domain setting yields a peak overall SECS score of 75.68, which is a +1.05 improvement over the best-performing unimodal setup. This enhancement generalizes across data partitions, with seen speaker performance increasing from 77.76 to 79.96, and unseen from 70.27 to 71.39. These findings demonstrate the efficacy of multi-domain, multi-attribute conditioning in reinforcing speaker identity consistency in the generated speech.\footnote{Speaker fidelity scores according to each specific category can be found in Appendix \ref{sec:attribute_classification}.}

\subsection{Compositional Analysis}
We conduct ablation studies to verify the components of the proposed model in Table \ref{tab:ablation2}. Eliminating audio-based attribute enhancement results in a noticeable drop of 1.05 points overall, while the further removal of visual enhancement leads to a larger decrease of 3.09 points, which highlights the greater impact of visual attribute alignment on preserving speaker identity. Moreover, we implement a separate experiment that employs just one vanilla transformer and takes the input speaker image as is, without any patch segmentation or feature aggregation (FGA). This results in a 6.26-point drop, which reaffirms the importance of conducting effective initial feature extraction of the input image. Finally, removing the data augmentation (AUG) strategy results in the lowest overall SECS score (63.23), highlighting the necessity of dataset augmentation for achieving robust generalization and speaker consistency. Taken together, these results demonstrate the synergistic effect of each architectural component in optimizing speaker similarity.

\begin{table}[t]
\centering
\resizebox{0.95\columnwidth}{!}{%
\begin{tabular}{lccc}
\toprule[1.2pt]
& \textbf{Seen SECS ($\uparrow$)} & \textbf{Unseen SECS ($\uparrow$)} & \textbf{Overall} \\ \midrule[1pt]
\rowcolor[gray]{0.9} \textbf{Ours}   & 79.96 & 71.39 & 75.68 \\ 
\begin{tabular}[c]{@{}l@{}}\,\; - Aud. Enhancement \end{tabular} & 77.76 & 70.27 & 74.63\\ 
\begin{tabular}[c]{@{}l@{}}\,\;\,\; - Vis. Enhancement \end{tabular} & 74.73 & 68.35 & 71.54 \\
\,\;\,\;\,\; - FGA   & 66.25 & 64.31 & 65.28 \\ 
\,\;\,\;\,\;\,\; - AUG   & 64.43 & 62.04 & 63.23 \\
\bottomrule[1.2pt]
\end{tabular}%
}
\caption{Ablation study results showing the effect of removing each individual component.}
\label{tab:ablation2}
\end{table}

%% file: conclusion.tex
\section{Conclusion}
In this paper, we have presented a novel end-to-end framework for FTV synthesis that emphasizes effective facial representation learning and cross-modal attribute alignment. In contrast to prior methods that depend on multi-stage pipelines or external pretrained models, our approach progressively aggregates local facial features to construct a robust identity embedding and then enforces semantic consistency through bilateral supervision of demographic attributes such as gender and ethnicity. Additionally, by incorporating a multi-view face-audio alignment strategy, we improve the model's ability to maintain vocal consistency across diverse visual inputs. Comprehensive evaluations across five assessment tasks and multiple objective metrics confirm that our method significantly improves speaker fidelity and identity preservation, thus offering a viable solution for personalized speech synthesis even in the absence of voice recordings.

%% file: limitation.tex
\section*{Limitations}
While our framework advances the generation of identity-consistent speech from facial images, it currently focuses on replicating overall vocal timbre. As such, it does not incorporate facial expressions, which are closely tied to emotional prosody. In future research, we plan to integrate facial expression information—potentially through emotion tagging—to enable expressive and emotionally congruent speech synthesis.
Furthermore, although our model includes demographic attributes such as gender and ethnicity via bilateral supervision, age-related features were excluded due to the highly skewed age distribution in the available dataset. Addressing this data imbalance and exploring age-aligned voice synthesis remains an important direction for future work. More broadly, a systematic investigation into the full range of facial attributes that influence perceived vocal characteristics will further enrich personalized and realistic voice synthesis.

\section*{Ethical Considerations}
The primary objective of this work is to develop a framework capable of synthesizing realistic, identity-consistent speech from facial images. This technology holds promising applications, particularly in assistive communication for individuals with speech impairments or those who have lost vocal capabilities due to neurological or physiological conditions. By enabling the generation of personalized voices, our framework may contribute to restoring a sense of vocal identity and enhancing self-expression for such individuals.
Nevertheless, FTV synthesis inherently involves the replication of personal identity cues, and thus raises potential ethical concerns. In particular, the potential for misuse, including unauthorized voice cloning, biometric spoofing, and identity impersonation, poses risks to personal privacy, consent, and digital security. These risks could lead to harmful applications such as fraud, misinformation, or defamation. We acknowledge these risks and emphasize the importance of responsible deployment and strict consent-based usage of FTV systems. Future work should incorporate robust safeguards such as imperceptible watermarking of synthetic speech, which may help to detect and deter misuse of generated voices. 

Furthermore, our model incorporates commonly defined demographic attributes such as gender and race during training for improved speaker identity modeling, which was informed by prior studies that indicate their perceptual relevance in voice characteristics. While these features were derived from publicly available datasets and used solely for the purpose of enhancing face-vocal alignment, we recognize the potential sensitivity of such demographic variables. 
We remain committed to conducting ethical and inclusive research and advocate for ongoing interdisciplinary dialogue between machine learning researchers, ethicists, and affected communities to ensure the responsible development of FTV technologies.

\section*{Acknowledgements}
This work was supported by the IITP(Institute of Information \& Communications Technology Planning \& Evaluation)-ITRC(Information Technology Research Center) grant funded by the Korea government(Ministry of Science and ICT)(IITP-2025-RS-2024-00437866)(47.5\%), Smart HealthCare for Police Officers Program(www.kipot.or.kr) through the Korea Institutes of Police Technology(KIPoT) funded by the Korean National Police Agency(KNPA, Korea)(No. RS-2022-PT000186)(47.5\%), and Institute of Information \& communications Technology Planning \& Evaluation (IITP) grant funded by the Korea government(MSIT) (No.RS-2019-II191906, Artificial Intelligence Graduate School Program(POSTECH))(5\%).